# Syntropic spin alignment at the interface between ferromagnetic and superconducting nitrides

**Authors:**

Qiao Jin,[1] Qinghua Zhang,[1] Bai He,[2] Yuting Zou,[1,3] Yonglong Ga,[4] Shengru Chen,[1,3] Haitao Hong,[1,3] Ting Cui,[1,3] Dongke Rong,[1] Jia-Ou Wang,[5] Can Wang,[1,2,6] Yanwei Cao,[4] Lin Gu,[7] Shanmin Wang,[8] Kun Jiang,[1] Zhi-Gang Cheng,[1,6] Tao Zhu,[1,2,6] Hongxin Yang,[9] Kui-juan Jin,[1,3,6,*] and Er-Jia Guo[1,3,*]

**Affiliations:**

[1] Beijing National Laboratory for Condensed Matter Physics and Institute of Physics, Chinese Academy of Sciences, Beijing 100190, China

[2] Spallation Neutron Source Science Center, Dongguan 523803, China

[3] Department of Physics & Center of Materials Science and Optoelectronics Engineering, University of Chinese Academy of Sciences, Beijing 100049, China

[4] Ningbo Institute of Materials Technology & Engineering, Chinese Academy of Sciences, Ningbo 315201, China

[5] Institute of High Energy Physics, Chinese Academy of Sciences, Beijing 100049, China

[6] Songshan Lake Materials Laboratory, Dongguan, Guangdong 523808, China

[7] National Center for Electron Microscopy in Beijing and School of Materials Science and Engineering, Tsinghua University, Beijing 100084, China

[8] Department of Physics, Southern University of Science and Technology, Shenzhen 518055, China

[9] National Laboratory of Solid-State Microstructures, School of Physics, Collaborative Innovation Center of Advanced Microstructures, Nanjing University, Nanjing 210093, China

E-mail: kjjin@iphy.ac.cn and ejguo@iphy.ac.cn

**Abstract**

The magnetic correlations at the superconductor/ferromagnet (S/F) interfaces play a crucial role in realizing dissipation-less spin-based logic and memory technologies, such as triplet-supercurrent spin-valves and "π" Josephson junctions. Here we report the coexistence of an induced large magnetic moment and a cryptoferromagnetic state at high-quality nitride S/F interfaces. Using polarized neutron reflectometry and *d. c.* SQUID measurements, we quantitatively determined the magnetization profile of S/F bilayer and confirmed the induced magnetic moment in the adjacent superconductor only exists below $T_C$. Interestingly, the direction of the induced moment in the superconductors was unexpectedly parallel to that in the ferromagnet, which contrasts with earlier findings in S/F heterostructures based on metals or oxides. The first-principles calculations verify the observed unusual interfacial spin texture is caused by the Heisenberg direct exchange coupling through *d* orbital overlapping and severe charge transfer across the interfaces. Our work establishes an incisive experimental probe for understanding the magnetic proximity behavior at S/F interfaces and provides a prototype epitaxial "building block" for superconducting spintronics.

**Teaser**: The superconducting nitrides exhibited an unexpectedly large magnetic moment and cryptoferromagnetism when brought into proximity with a strong ferromagnet.

**Introduction**

The interface between a superconductor (S) and a ferromagnet (F) is a topic of ongoing research in condensed matter physics. The interaction between superconductivity and ferromagnetism leads to fascinating phenomena, including (inverse) magnetic proximity effects [1-5], spin-triplet superconductivity [6,7], and the emergence of Majorana fermions [8]. Of particular interest is the magnetic proximity effect, which directly reflects the exchange interaction between the spins of electrons across S/F interfaces, resulting in the suppression of magnetic order or the appearance of unconventional superconductivity [9,10]. When a magnetic material is proximity to a superconductor, the magnetic order parameter can penetrate the superconductors over a short distance (usually a few nanometers) [11]. This causes spatial variation of the superconducting (SC) order parameters near the interface and disrupts the Cooper pairs, significantly impacting the macroscopic physical properties of materials on both sides. In addition to the fundamental understanding based on S/F interfaces, extensive research has been conducted on the coupling between superconducting condensate states and magnetic-exchange spin coupling. This research has opened up a new field of study known as superconducting spintronics, which aims to develop dissipation-less spin-based devices, including memories [12], supercurrent spin-valves [13-15], and "π" Josephson junctions [16]. Therefore, it is particularly relevant to experimentally verify the creation of a spin-polarized superconducting state with parallel or anti-parallel spin alignment at proximity-engineered S/F interfaces, as this is a crucial step in advancing this research direction.

So far, the genuine mechanism behind the magnetic proximity effect at S/F interfaces is still debated. Earlier research on S/F heterostructures composed of metal alloys reported an oscillation of the superconducting transition temperature as a function of the ferromagnetic layer thickness, demonstrating the possible unconventional propagation of superconducting pair waves in the system due to the strong exchange field [17-21]. With the development of advanced thin-film synthesis techniques, increasing attention has been focused on high-quality epitaxial S/F interfaces between a high-temperature superconductor ($YBa_2Cu_3O_7$) and a fully spin-polarized half-metal ferromagnet ($La_{1-x}Ca_xMnO_3$). [22-25]. Evidence from both X-ray magnetic circular dichroism (XMCD) [26] and polarized neutron reflectometry (PNR) [27-30] confirm a reduction of the magnetic moment at S/F heterointerfaces and the antiparallel spin alignment between Cu and Mn ions in each component. The magnetic proximity effects in $YBa_2Cu_3O_7/La_{1-x}Ca_xMnO_3$ interfaces are rather complex and depend on many impact factors, such as the electronic state of magnetic layers [26], the thickness of SC layer [29], nonhomogeneous domain structures [23-24], etc. Although the extensive studies had been conducted in both metals and oxides, the opposite magnetic coupling through the S/F interfaces makes it difficult to control the spin collinearity by external magnetic fields. Furthermore, the experimental preliminary confirmation of the existence of the triplet component in S/F structures or the formation of cryptoferromagnetic state in SC had seldomly be explored. The active and reversible control of triplet supercurrents in spin-valve structure can be highly advantageous for developing superconducting spintronics in future.

In this work, we investigated transition metal nitride bilayers composed of the superconducting VN and the ferromagnetic $Fe_3N$ using conventional electrical and magnetic transport measurements as well as PNR technique. Our findings indicate that the proximity of the SC layers to $Fe_3N$ can significantly reduce the SC order parameters and result in unconventional magnetic behaviors. We also confirmed that an induced large magnetic moment extends a few nanometers into the SC layers, with its sign unexpectedly parallel to the magnetic moment in the $Fe_3N$ layer, which contradicts earlier reports and theoretical predictions. Instead of superexchange coupling at oxide interfaces, the Heisenberg direct exchange coupling at nitride interfaces is further verified by first-principles calculations. This study provides a detailed microscopic picture that offers intriguing insights into the magnetic correlations at S/F interfaces.

**Results**

The $Fe_3N$/VN bilayer, as well as $Fe_3N$ and VN single layers, were grown on (00$l$)-oriented $Al_2O_3$ substrates by pulsed laser deposition (PLD) assisted by RF nitrogen plasma (for experimental details, see Methods) [31-34]. Further details on the structural characterizations of the $Fe_3N$ and VN single layers can be found in our previous works [34,35]. Figure 1a shows an XRD $\theta$-$2\theta$ scan of the $Fe_3N$/VN bilayer, which exhibits narrow diffraction peaks with high-order Laue thickness fringes, indicating the epitaxial growth of a high-quality bilayer. The VN layer has a (111) orientation grown on $Al_2O_3$ substrates, while the $Fe_3N$ layer, grown on (111)-oriented VN, has a (100) orientation. Notably, the structural orientation of $Fe_3N$ in the bilayer is differs from that of direct growth of $Fe_3N$ single films on $Al_2O_3$ [35]. The insets of Figure 1a illustrate the atomic arrangements of the related nitrides and substrates. The epitaxial relationship between two nitride layers is enabled by their compatible crystallographic symmetry and relatively small misfit strain. XRR measurements were used to examine the well-defined interfaces and chemical compositions within the bilayer. The root-mean-square roughness of the $Fe_3N$/VN interface, averaged over the coherence of the X-ray beam projected on the sample's surface ($\approx$ tens of millimeter$^2$), was found to be 5.2 ± 0.5 Å. Figure 1b shows the chemical depth profile obtained from fitting a model to XRR experimental data. The X-ray scattering length density (SLD) profile revealed that the electron density at the surface of $Fe_3N$ is significantly lower compared to the interior part of the $Fe_3N$ layer. Additionally, the SLDs within the VN layer was nonuniform, with the SLD of the VN interfacial layer, having a thickness of 4.6 nm, greater by ~5% compared to that of the rest of the VN layer.

We performed STEM measurements on the identical $Fe_3N/VN$ bilayer. Figure 1c indicates the high crystallinity of both $Fe_3N$ and VN layers, and the bilayer appears to be free of noticeable defects. The $Fe_3N$ surface layer has a relatively lower HAADF brightness compared to the rest of the layer, consistent with XRR fitting results. Further investigation of the chemical composition was carried out using elemental-specific electron energy loss spectroscopy (EELS), as shown in Figure 1d and Supplementary Figures S1 and S2. We discovered that the $Fe_3N$ surface layer contains a significant amount of oxygen, indicating the oxidization of Fe ions. High-magnification STEM imaging was taken at the surface region of the $Fe_3N/VN$ bilayer (Supplementary Figure S3), revealing that the top surface has a spinel ($Fe_3O_4$) structure with clearly visible octahedral and tetrahedral iron atomic columns. Therefore, the surface of the $Fe_3N/VN$ bilayer comprises a mixture of $Fe_3N$ and $Fe_3O_4$ single-crystalline films. The oxidization of the $Fe_3N$ layer is unavoidable phenomenon when exposed to ambient conditions, which is commonly observed in other iron nitride compounds [36]. Notably, the $Fe_3O_4$ surface layer does not affect the quality of the buried interfaces. The atomic-resolved interfaces of $Fe_3N/VN$ and $VN/Al_2O_3$ are shown in Figures 1e and 1f, respectively. Both interfaces are clearly abrupt, with well-aligned atoms and no significant intermixing inside the VN layers. Additionally, XAS measurements were performed on the $Fe_3N/VN$ bilayer at room temperature with a small incident angle (large portion of irradiated area). The resulting spectra were taken at both resonant N $K$- and Fe $L$-edges (Supplementary Figure S4), indicating that the $Fe_3N$ layers have sufficient nitrogen content, and the valence state of iron ions remains +3. In contrast to the severe charge transfer observed between Fe and Mn ions at the $YBa_2Cu_3O_7/La_{1-x}Ca_xMnO_3$ interfaces [24], no significant oxidization state changes were observed for both Fe and V ions at the $Fe_3N/VN$ interfaces.

Figure 2a presents the temperature-dependent resistivity of a VN single film and a $Fe_3N/VN$ bilayer at zero magnetic field. Prior to the sharp superconducting transition, all samples exhibit metallic phases. The normal state Hall resistivity measured at 10 K confirms that the charge carriers are electrons, and the carrier densities of the VN single film and $Fe_3N/VN$ bilayer fall within the range of $(19.1 – 26.3) \times 10^{22}$ cm$^{-3}$ (Table I). The superconducting transition temperature ($T_C$) of the VN single film is ~ 7.78 K and is suppressed by ~ 1.55 K after capping a ferromagnetic $Fe_3N$ layer. This effect is typically attributed to the presence of spin-polarized quasiparticles at the interface, which can enhance the scattering of the Cooper pairs responsible for superconductivity. The temperature-dependent resistivity of the two samples were recorded under different magnetic fields, both parallel and perpendicular to the sample's surface plane, up to 9 T (Supplementary Figure S5). Both samples exhibit conventional suppressed superconductivity with increasing field. The VN is a typical type-II superconductor with upper critical field ($H_{c2}$) [37]. Figures 2b and 2c show the in-plane and out-of-plane magnetic field-dependent resistivity of a $Fe_3N/VN$ bilayer, with temperature fixed at values from 3 to 8.5 K, respectively. The same transport measurements were conducted repeatedly on a VN single film (Figures 2d and 2e). The $H_{c2}$ is determined at fields when resistivity reduces to the half value of its normal state. In Figure 2f, we plotted the temperature dependence of the $H_{c2}$ for two samples when $H // ab$ and $H // c$. Notably, both samples exhibit a smaller $H_{c2}$ ($H // c$) compared to $H_{c2}$ ($H // ab$). These data can be well-fitted by a single-band Werthamer-Herlfand-Hohenberg (WHH) equation (dashed lines) [38]. The fitting parameters for the two samples are summarized in Table I. The mean free path of the VN layer increases by ~ 30% after proximity to a $Fe_3N$ layer. Simultaneously, the estimated coherence length increases from ~405 nm (VN) to ~422 nm ($Fe_3N/VN$).

Earlier studies have shown that the ferromagnetism in a F layer can be influenced by superconductivity when the thickness of the F layer is much smaller than the coherence length of the SC layer [2,3]. To text this hypothesis, we measured the field-dependent magnetization ($M$) of a $Fe_3N$ single layer and a $Fe_3N/VN$ bilayer (Figures 3a and 3b and Supplementary Figure S6). We recorded both in-plane and out-of-plane $M$ at fixed temperatures across $T_C$. The magnetic easy-axis is along the in-plane direction, consistent with the magnetic behavior of 2D thin films [35]. The total $M$ of both samples was normalized to the thickness of the $Fe_3N$ layer. At the high field regions, $M$ of both samples is saturated. Interestingly, we observed that the $M_{sat}$ of a $Fe_3N/VN$ bilayer is lower than that of a $Fe_3N$ single layer below $T_C$, and these values returned to the same level as that of a $Fe_3N$ single layer when the temperature is above $T_C$. We believe that the exchange interactions between the superconducting condensate and the magnetic order parameter reduce the energy of the bilayer system, possibly leading to a cryptoferromagnetic state [39]. When the temperature is below $T_C$, the VN becomes superconducting, and some of the electrons located in the $Fe_3N$ layers may condensate into Cooper pair, reducing the absolute value of the total magnetization in the $Fe_3N/VN$ bilayer, as they do not contribute to the total magnetization.

In additional to the change in the total magnetization of a $Fe_3N/VN$ bilayer, the magnetic state of the superconducting VN layer can be correspondingly modified due to the magnetic proximity effect [1-5]. The magnetization measurements on the $Fe_3N/VN$ bilayer confirm the presence of antimagnetic signals when the VN layer enters the SC state (Figure 3c). The $T_C$ obtained from the $M$-$T$ curves is consistent with the electrical transport measurements (inset of Figure 3c). From the $M$-$H$ measurements, we observe abnormal behavior in the hysteresis loops of the $Fe_3N/VN$ bilayer in the low field region when the temperature is lowered below $T_C$. When $H // ab$, the magnetic moment initially increases with applied fields and then returns to a constant value of $M_{sat}$ beyond a critical field ($H_{SC}$). This behavior suggests the presence of an induced magnetic moment in the VN layer for temperatures below $T_C$. Simultaneously, the out-of-plane magnetic moment of the Fe$_3$N/VN bilayer jumps to incredible values at the small fields and exhibits a greatly increased coercive fields ($H_C$) when the VN layer is in the superconducting state. Figures 3d and 3e show $H_{SC}$ and $H_C$ as a function of the temperature, respectively. Clearly, $H_{SC}$ and $H_C$ in the SC state are orders of magnitudes larger than those in the normal state and reduce progressively as the temperature increases. When the temperature is above $T_C$, the anomaly in the magnetization along both in-plane and out-of-plane direction disappears.

To quantitatively determine the interfacial magnetization profile, we perform systematic PNR measurements on a Fe$_3$N/VN bilayer, as shown in Figure 4a. The specular reflectivity ($R$) of the bilayer was measured as a function of the wave vector transfer ($q$), with $R^+$ and $R^-$ representing the reflectivities for neutrons with spins parallel or anti-parallel to the applied magnetic field, respectively. The PNR measurements were performed under a magnetic field of 1 T at 3.5 K, when the Fe$_3$N/VN bilayer was in the SC state. Figure 4b shows the PNR data with the reflectivity normalized to the asymptotic value of the Fresnel reflectivity $R_F$ (= $16\pi^2/q^4$). The large $q$-dependent splitting between the two reflectivities reflects the large net magnetization of the Fe$_3$N/VN bilayer. To fit the PNR data, we used the chemical depth profile obtained from XRR fitting to strictly constrain a model for PNR fitting. We obtained the nuclear and magnetic SLD profiles corresponding to the chemical and magnetization distribution as a function of film thickness, as shown in Figures 4e and 4f. The best fits demonstrate that the Fe$_3$N layer has a magnetization of 1630.9 ± 7.1 kA/m and the Fe$_3$O$_4$ surface layer has a much lower magnetization of 339.5 ± 6.3 kA/m. The magnetization of Fe$_3$N interior layer is significantly larger than the magnetization measured by SQUID, which underestimated the value by assuming a uniformed distribution of magnetization across the entire Fe$_3$N layer. Additionally, we observed that the VN interfacial layer in close proximity to Fe$_3$N exhibits a significant net moment of 115.6 ± 2.5 kA/m ,which aligns parallel to the Fe$_3$N magnetization. The spin texture schematic at the Fe$_3$N/VN interfaces is depicted in Figure 4a. Confidence in this interpretation of syntropic spin alignment at the interface is further reinforced by fitting the spin asymmetry (SA) curve derived from experimental data (Figure 4c). To fit the PNR-SA data, we considered three different scenarios: (1) parallel spins (positive $M$); (2) no spins (zero $M$); and (3) antiparallel spins (negative $M$) of the VN interfacial layer with respect to the spins of Fe$_3$N layer (Figure 4c and Supplementary Figure S7). The best fit to the PNR data, with a lowest $\chi^2$ metric ~ 1.1, demonstrates that the ferromagnetic coupling across the interface is an intrinsic nature. Under the same magnetic field of 1 T, we conducted PNR measurements on the Fe$_3$N/VN bilayer at 15 K (Figure 4d). The induced magnetic moment in the VN interfacial layer decreases to zero (Figure 4f), which is consistent with SQUID results.

To confirm the observation of the induced moment in VN interfacial layer, we performed further control measurements on the Fe$_3$N/VN bilayer at a fixed temperature of 6.5 K with varying magnetic fields between 2 kOe and 1 T (Supplementary Figure S8). The measuring temperature was carefully chosen at the boundary of the phase transition between the SC state and normal state. Under a magnetic field of 2 kOe, the Fe$_3$N/VN bilayer remained in the SC state. The VN interfacial layer exhibits a net magnetic moment of 96.5 ± 3.5 kA/m (Figure 4g). As the magnetic field was increased to 1 T, the magnetization of the Fe$_3$N layer increased only by ~ 5% because the Curie temperature of ferromagnetic Fe$_3$N is far beyond room temperature. However, no induced moment in the VN interfacial layer was observed. This was because the SC state of the bilayer was destroyed, and it enters the normal state at high magnetic fields. Therefore, our experimental results provide clear evidence that a parallel magnetic moment of in the VN layer close to the interface only exists once the bilayer enters the SC state. At this moment, we could not argue the formation of magnetic domains below $T_C$ due to the presence of magnetic vortices creating electrodynamic forces [23, 40]. The off-specular PNR experiments did not observe the clean Bragg peak diffraction below $T_C$. At least, our results clearly rule out the existence of a triplet supercurrent at the interface that previously had been proposed to explain the observed magnetism at S/F interfaces [41-43], otherwise the induced moment of VN interfacial layer should have an opposite sign with respect to that of Fe$_3$N.

Furthermore, earlier work demonstrates the magnetic exchange coupling in the YBa$_2$Cu$_3$O$_7$/La$_{1-x}$Ca$_x$MnO$_3$ heterostructures strongly depends on the thickness of superconducting layer [23, 26, 44]. We measured the transport properties of a VN single film and a Fe$_3$N/VN bilayer with a VN layer thickness of ~ 6 nm (Figure 5a). The 6-nm-thick VN film exhibits a superconducting behavior with $T_C$ ~ 4.5 K. The superconducting order parameters show an increased anisotropy compared to those of a thick VN single layer (Supplementary Figure S9). Moreover, the Fe$_3$N/VN bilayer maintains the ferromagnetic metallic phase with clear anisotropic MR when temperature is lowered down to 300 mK (Figure 5b and Supplementary Figure S10). These results suggest the strong suppression of superconducting ordering by a ferromagnetic capping layer. We note that the induced ferromagnetism and superconductivity cannot coexist in the interfacial VN layers.

Previous experimental evidence of long-range penetrated ferromagnetism into conventional superconducting oxides reveals an antiparallel magnetic coupling with respect to the ferromagnetic layers. [23-30]. In those cases, the superexchange coupling between 3$d$ transitional metal elements via oxygen 2$p$ orbitals results in an antiferromagnetic spin alignment. However, our findings demonstrate that spins can be aligned parallel to each other across S/F interfaces (Figure 5c). We performed the first-principles calculations on Fe$_3$N/VN interfaces based on density functional theory (DFT). Both FM and AFM magnetic configurations are initially texted (Supplementary Figure S11).

We calculated Heisenberg exchange coupling constant ($J$) between Fe and V atoms near the interface is 4.28 meV. The positive sign indicates the ferromagnetic coupling across the interfaces. Besides, we find that the electron clouds (marked in yellow) around the interfaces is denser than those in the interior parts of VN and Fe$_3$N layers (Figure 5d). The calculation results indicate the electron transfer from Fe ions to V ions across the interface due to the charge density difference. We further calculated the contour density for the spin up (red) and spin down (yellow) components in Figure 5e. We found that the spin polarization exhibits a large positive value for the constructed heterointerface. The density of states (DOS) for spin-up channels are significantly larger than those of spin-down channels at the Fermi level, implying that the heterointerfaces exhibit the ferromagnetic coupling.

The ferromagnetically spin alignment between Fe and V ions is in sharp contrast to the earlier works on the spin coupling at oxide interfaces. We propose a qualitative physical explanation for the observed magnetic behaviors at the Fe$_3$N/VN interfaces. When the temperature falls below $T_C$, magnetic vortices form in the superconducting VN layer, but they do not influence the in-plane magnetization of the bilayer (Figure 3f). In the Fe$_3$N/VN heterostructures with strong exchange fields, the number of spin-up and spin-down electrons is imbalanced, leading to a magnetic "*leakage*" from the Fe$_3$N layer to into the VN layer and electron polarization in the VN layer persists over a short length scale. In this scenario, the direction of the induced magnetic moment in the VN interfacial layer aligns parallel to that in the Fe$_3$N layer. This finding is supported by numerical solutions to the Bogoliubov-de Gennes equations [2, 3], which are consistent with our SQUID and PNR results. However, when the magnetic field switches to the out-of-plane direction (Figure 3g), a different situation arises. At small fields, the Fe$_3$N/VN bilayer remains in SC state, and the magnetic flux only penetrates through the vortices, resulting in ultra-large effective magnetic fields. This leads to the observation of incredibly large magnetic moments in the Fe$_3$N/VN bilayer at small magnetic fields. Additionally, the strong competition between ferromagnetic and superconducting order parameters leads to enhanced coercive fields when the VN in the SC state. When the magnetic field exceeds $H_{c2}$ or the temperature increases above $T_C$, the Fe$_3$N/VN bilayer enters the normal state, the impact from the superconducting VN layer vanishes, greatly weakening the strength of interfacial magnetic coupling. The $M_{sat}$ and $H_C$ of the Fe$_3$N/VN bilayer return to the values that are nearly identical to those of a Fe$_3$N single layer.

Finally, we performed repeated measurements on a high-quality Fe$_3$N/TiN bilayer and compared the results to those obtained from a TiN single layer (Supplementary Figures S12-S14). Similar effects, such as the large anisotropy of $H_{c2}$ and the suppression of $T_C$, were observed in the Fe$_3$N/TiN bilayer. Before TiN enter the SC state, we observe the coexistence of magnetoresistance and the opening of a superconducting pseudogap, suggesting the presence of strong competing orders at the critical transition temperature. Due to the significant suppression of $T_C$ to ~ 1.28 K, systematic magnetometry and PNR measurements would be very challenge. Testing whether the induced moment in TiN remains parallel to that of ferromagnetic $Fe_3N$ will be examined in future studies.

**Discussions and conclusions**

In summary, our work presents new findings regarding the induced magnetic moment in a superconducting VN layer in proximity to a strong ferromagnetic $Fe_3N$ layer. Our results differ from previous studies, as we have discovered that the spin orientation in VN is ferromagnetically coupled to the magnetization of the adjacent $Fe_3N$ layer, revealing a unique form of magnetic coupling across the S/F interface. Furthermore, we propose the possibility of cryptoferromagnetic states in the S/F bilayers, where superconductivity can persist despite the presence of a ferromagnetic background. Our research sheds light on the potential for exploring complex competing orders in the large family of transition metal nitrides, such as ZrN and NbN [45,46], which are known to host superconducting transitions. Such materials have not been thoroughly investigated in the past, and our study provides a strategy for future exploration. Additionally, we recommend prioritizing the testing of binary or perovskite-type nitrides containing 5*d* elements [47,48]. These materials not only have competing phases, but also exhibit a strong spin-orbital (SO) interaction. This interaction has an energy scale comparable to the superconducting gap, which may have a significant impact on the penetration length of the triplet component into the superconductor, thereby dominating the long-range proximity effect in S/F structures.

**Methods**

**Sample synthesis**

The VN and $Fe_3N$ thin films were fabricated on (001)-oriented single-crystalline $Al_2O_3$ substrates by pulsed laser deposition (PLD). Nearly stoichiometric binary nitride targets were synthesized using a high-pressure reaction route at the high-pressure Laboratory of South University of Science and Technology (SUSTech). The VN films were deposited at a substrate temperature of 750 ºC, while the $Fe_3N$ films were deposited at a lower substrate temperature of 300 ºC to maintain their high crystallinity. During deposition, the density of laser fluence was maintained at ~ 1 J/cm$^2$, and the base pressure was kept at around $10^{-8}$ Torr. A radio frequency (RF) plasma source with tunable input power (100-400 W) and partial pressure of $N_2$ gas ($10^{-3}$-$10^{-6}$ Torr) was applied to generate highly active nitrogen atoms. The generated nitrogen plasma helped to compensate the nitrogen vacancies in the films, enabling the as-grown films to maintain their intrinsic characteristics. The thickness of VN ($Fe_3N$) films was controlled by counting laser pulses. The TiN films were fabricated using a home-made RF magnetron sputtering system at the Ningbo Institute of Materials Technology and Engineering, CAS. The base pressure of the sputtering chamber is below $3\times10^{-8}$ Torr. The TiN films were deposited at 800 °C under a pure nitrogen pressure of 0.02–0.03 Torr. The RF generator power was maintained at 100 W during the deposition. The thickness of the TiN films was controlled by counting sputtering time. Subsequently, the TiN films were capped with a $Fe_3N$ films using PLD.

**Structural and electronic characterizations**

X-ray reflectivity (XRR) measurements were conducted to confirm the structural integrity, layer thickness, interface/surface roughness, and densities of all layers, from which we could obtain the chemical profiles of $Fe_3N$/VN and TiN/VN bilayers. The θ-2θ scans were performed on a high-resolution four-circle X-ray diffractometer (Panalytical MRD X'Pert 3) with Cu $K\alpha$ radiation. The synchrotron based XRD measurements were performed at the beamline 1W1A of the Beijing Synchrotron Radiation Facility (BSRF). The wavelength of synchrotron X-ray is 1.24 Å. The microstructures of a $Fe_3N$/VN bilayer were examined using JEM ARM 200CF electron microscopy at the Institute of Physics, Chinese Academy of Sciences. The samples were prepared using $Ga^+$ ion milling after the mechanical thinning. Both HAADF and ABF imaging were performed in the scanning mode. Elemental-specific EELS mapping were performed at the O $K$-, N $K$-, V $L$-, Fe $L$-edges from the interested regions. X-ray absorption spectroscopy (XAS) measurements were performed on a $Fe_3N$/VN bilayer using the total-electron yield (TEY) method. All structural characterizations were performed at room-temperature.

**Transport and magnetic characterizations**

Transport properties of VN and TiN single films and $Fe_3N$/VN and $Fe_3N$/TiN bilayers were measured using standard van der Pauw geometry. The temperature and magnetic field dependent resistivity measurements were conducted using a 9T–PPMS. The *ac* current was kept at a minimum requirement of 1 μA to avoid the Joule heating. The magnetic properties of a $Fe_3N$ single film and a $Fe_3N$/VN bilayer were characterized by a SQUID equipped with high-temperature unit. Both in-plane and out-of-plane magnetization were obtained at the variable temperatures and fields. For the antimagnetic measurements of a $Fe_3N$/VN bilayer, the sample was zero-field cooled and measured during sample warming up process at a field of 200 Oe.

**PNR measurements**

PNR measurements were carried out at MR beamline of Chinese Spallation Neutron Source (CSNS), Dongguan, Guangdong Province. The size of a $Fe_3N$/VN bilayer is $10 \times 10 \times 0.5$ mm$^3$. We performed PNR measurements at fixed temperatures of 3.5 and 15 K under an in-plane magnetic field of 1 T. These temperatures were chosen because the VN films stay in the superconducting (3.5 K) and normal state (15 K) during the PNR measurements. Additionally, we conducted a control measurement by fixing the temperature at 6.5 K. Similarly, the VN films can be switched between superconducting and normal state under magnetic fields of 2 kOe and 1 T, respectively. The specular neutron reflectivity ($R$) of a Fe$_3$N/VN bilayer were recorded as a function of the wave vector transfer q (= $4\pi\sin\alpha/\lambda$), where α is the incident angle of neutron beam and λ is the wavelength of neutrons. Both $R^+$ and $R^-$ were recorded when neutrons with spins parallel or antiparallel to the applied fields (corresponding to the spin-up and spin-down neutrons), respectively. We fitted PNR data to a model (including the layer thickness and chemical roughness) that was obtained by XRR fitting using the formalism of Parratt. In these cases, the nuclear scattering length densities (nSLD) for the Fe$_3$N and VN were fixed to their bulk values of ~ $8.9 \times 10^{-6}$ Å$^{-2}$ and ~ $5 \times 10^{-6}$ Å$^{-2}$, respectively.

**First-principles calculations**

The first-principles calculations within the framework of density-functional theory (DFT) based on Vienna *ab* initio Simulations Package (VASP) [49]. Projected augmented wave (PAW) and generalized gradient approximation (GGA) methods are adopted to treat the valence electrons-ion and exchange-correction effects, respectively [50]. In order to correctly describe the 3*d* electrons, the GGA+U method is employed with an effect Hubbard U value $U_{Fe}$ = 1.5 eV for Fe element and $U_V$ = 3 eV for V element [51]. Otherwise, a 500 eV cutoff energy for plane-wave expansion and 6 × 6 × 1 *k*-point meshes are set for determining the most stable structure. Meantime, the Fe$_3$N/VN heterointerface is relaxed until the Hellmann-Feynman force acting on each atom become smaller than $1\times10^{-3}$ eV/Å. Finally, the Heisenberg exchange coupling constant (*J*) between interfacial Fe and V elements is solved via comparing the energy between ferromagnetic (FM) and antiferromagnetic (AFM) configurations (as shown in a 1×√3×1 supercell is constructed, Supporting Information Fig. S11).

**Acknowledgements**

We sincerely thank the fruitful discussions with Zhiqi Liu (BUAA), Guoqiang Yu (IOP-CAS), and Enke Liu (IOP-CAS). The magnetization measurements were performed by the generous help from Tengyu Guo (Songshan Lake Mater. Lab.). **Funding**: This work was supported by the National Key Basic Research Program of China (Grant Nos. 2019YFA0308500 and 2020YFA0309100), the National Natural Science Foundation of China (Grant Nos. U22A20263, 52250308, 11974390, 11721404), the Guangdong-Hong Kong-Macao Joint Laboratory for Neutron Scattering Science and Technology (HT-CSNS-DG-CD-0080/2021), and the Strategic Priority Research Program (B) of the Chinese Academy of Sciences (Grant No. XDB33030200). PNR experiments were conducted at the Beamline MR of Chinese Spallation Neutron Source, CAS. The synchrotron based XRD measurements were performed at the beamline 1W1A and XAS experiments were performed at the beamline 4B9B of the Beijing Synchrotron Radiation Facility (BSRF) via user proposals. **Author contributions: The nitride samples were grown by Q.J.; TEM lamellas were fabricated with FIB milling and TEM experiments were performed by Q.H.Z. and L.G.; PNR measurements were conducted by H.B. under the guidance of T.Z. XAS measurements were performed by Q.J., S.R.C., H.T.H., T.C., and J.O.W.; Q.J., and S.R.C., worked on the structural and magnetic measurements. Y.L.G. and H.X.Y. performed the first-principles calculations based on density functional theory. Y.T.Z. and Z.G.C. performed the ultra-low temperature transport measurements. C.W. participated the discussions and K.J.J. provided important suggestions during the manuscript preparation. E.J.G. initiated the research and supervised the work. Q. J. and E.J.G. wrote the manuscript with inputs from all authors. **Competing interests:** The authors declare that they have no competing financial interests. **Data and materials availability:** The data that support the findings of this study are available on the proper request from the first author (Q.J.) and the corresponding author (E.J.G.). **Additional information**: Supplementary information is available in the online version of the paper. Reprints and permissions information is available online.

**Table and table caption**

|  | VN | VN | Fe$_3$N/VN | TiN | Fe$_3$N/TiN |
|---|---|---|---|---|---|
| $t$ (nm) | 6.0(1) | 33(1) | 6.5(1)/33(1) | 55(1) | 6.5(1)/55(1) |
| $T_C$ (K) | 4.5(5) | 7.78(3) | 6.23(2) | 5.35(2) | 1.28(5) |
| $H_{C2}$ ($H // c$) (T) | 11.47(5) | 7.34(5) | 5.25(5) | 0.33(7) | 0.029(5) |
| $H_{C2}$ ($H // ab$) (T) | 2.95(6) | 8.74(3) | 8.01(5) | 1.05(5) | 0.23(1) |
| $n$ ($\times 10^{22}$ cm$^{-3}$) | 17.5(3) | 26.3(2) | 19.1(3) | 8.36(3) | 6.80(2) |
| $\kappa_F$ ($10^6$ m/s) | 2.0(4) | 2.29(2) | 2.09(2) | 1.56(3) | 1.46(2) |
| $L_C$ (nm) | 612(3) | 405(3) | 422(5) | 401(3) | 1568(8) |
| $\lambda$ (nm) | 0.86(2) | 1.03(5) | 1.32(4) | 11.85(8) | 10.21(5) |

**Table I. Summary of the parameters for the VN and TiN superconducting single layers, and Fe$_3$N/VN and Fe$_3$N/TiN bilayers**, including layer thickness ($t$), superconducting transition temperature ($T_C$), upper critical field ($H_{C2}$), carrier density ($n$), Fermi vector ($\kappa_F$), coherence length ($L_C$), and mean free path ($\lambda$). The electronic parameters, such as resistivities, carrier densities, and mobilities, of single layer and bilayers were extracted from transport measurements conducted in the normal state (at 10 K). Estimated errors are listed in parentheses.

**Figures and figure captions**

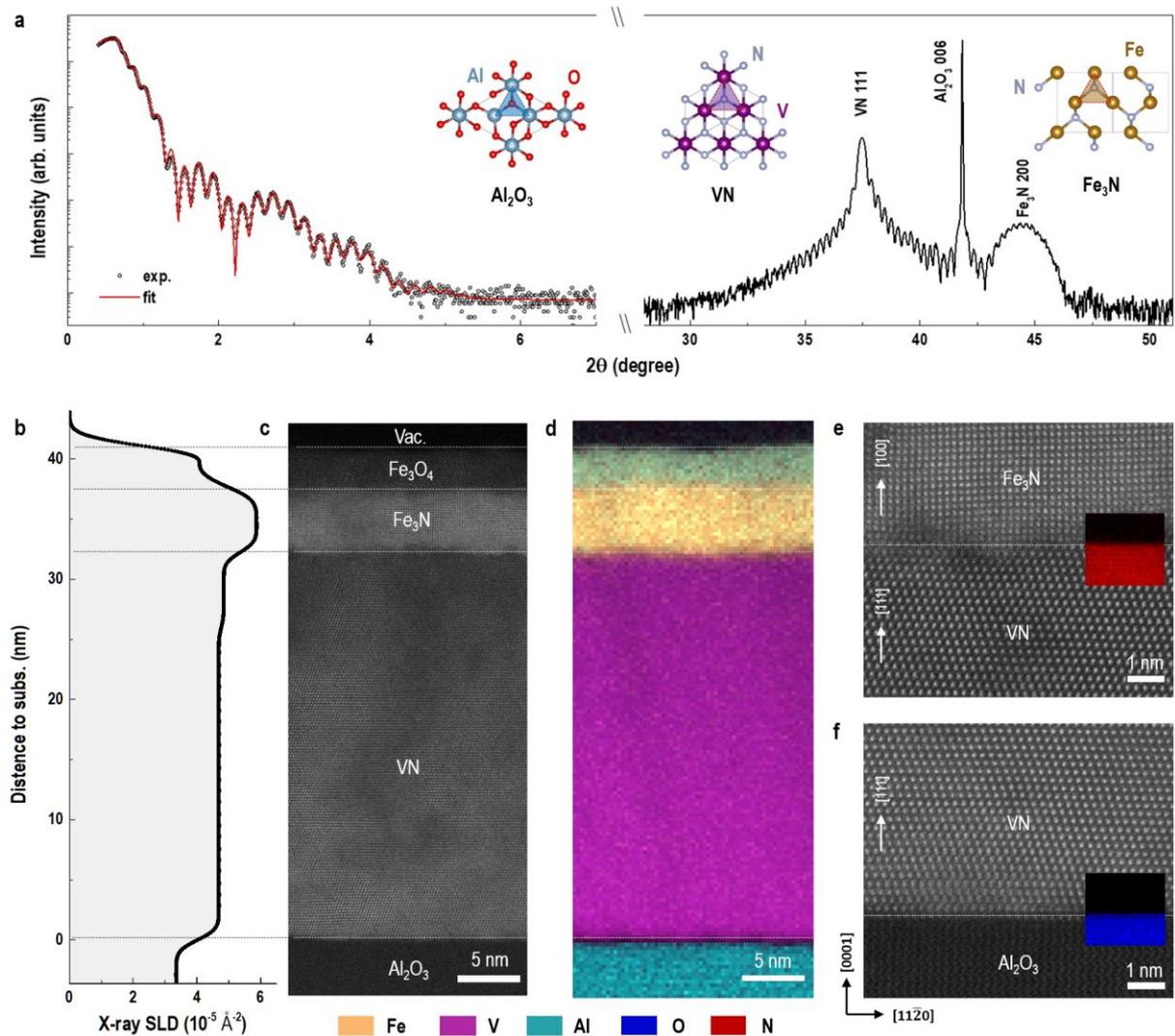

**Figure 1. Atomically sharp interface between ferromagnetic Fe₃N and superconducting VN.** (a) XRR and θ-2θ scan of an Fe₃N/VN bilayer grown on an Al₂O₃ substrate. The open symbols represent the experimental data, and the red line represents the best fit. The insets show the plane-view of crystal structures for all layers and substrates. (b) X-ray scattering length density (SLD) depth profile across the Fe₃N/VN heterointerfaces. (c) High-angle annular dark-field scanning transmission electron microscopy (HAADF-STEM) image, and (d) combined electron-energy loss spectroscopy (EELS) spectrum image collected from a Fe₃N/VN bilayer. The colors in the EELS mapping indicate the distribution of elements. Representative high-resolution TEM image at (e) Fe₃N/VN and (f) VN/Al₂O₃ interfaces are also shown. The colored insets in (e) and (f) show the integrated intensities of N $K$- and O $K$-edges, respectively. The nitrogen content in Fe₃N is comparably lower than that in VN.

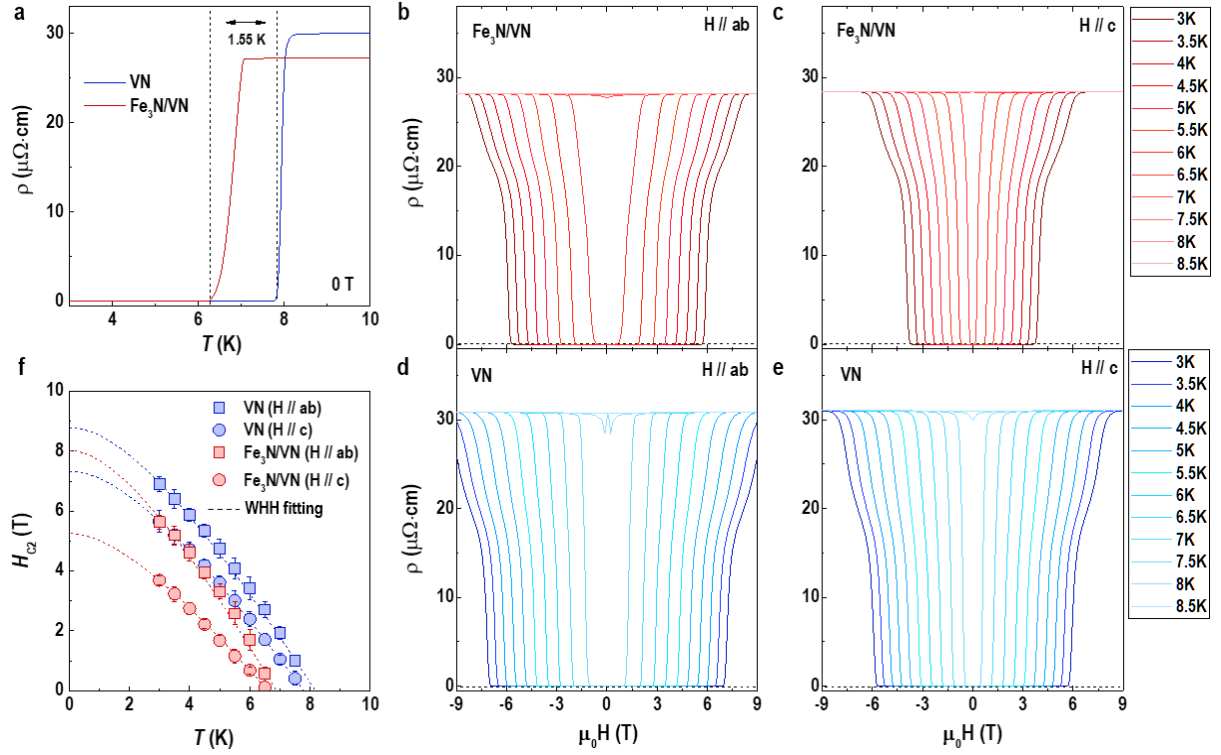

**Figure 2. Suppression of superconducting transition temperature ($T_C$) by a strong ferromagnet Fe$_3$N**. (a) Temperature dependent resistivity of a VN single layer and a Fe$_3$N/VN bilayer at zero magnetic field. The $T_C$ of the VN films was suppressed by ~1.55 K. (b) and (c) Magnetic-field dependent resistivity of the Fe$_3$N/VN bilayer at various temperatures when the magnetic field was applied parallel to the *ab*-plane and the *c*-axis, respectively. Similar measurements were performed on the VN single layer, as shown in (d) and (e). (f) Temperature dependence of the upper critical field ($H_{C2}$) for the VN single layer and Fe$_3$N/VN bilayer. The square and circle symbols represent $H_{C2}$ when the magnetic field was applied parallel to the *ab*-plane and the *c*-axis, respectively. The dashed lines are fitting curves based on the Werthamer-Helfand-Hohenberg (WHH) model. The details of fitting parameters are summarized in Table I.

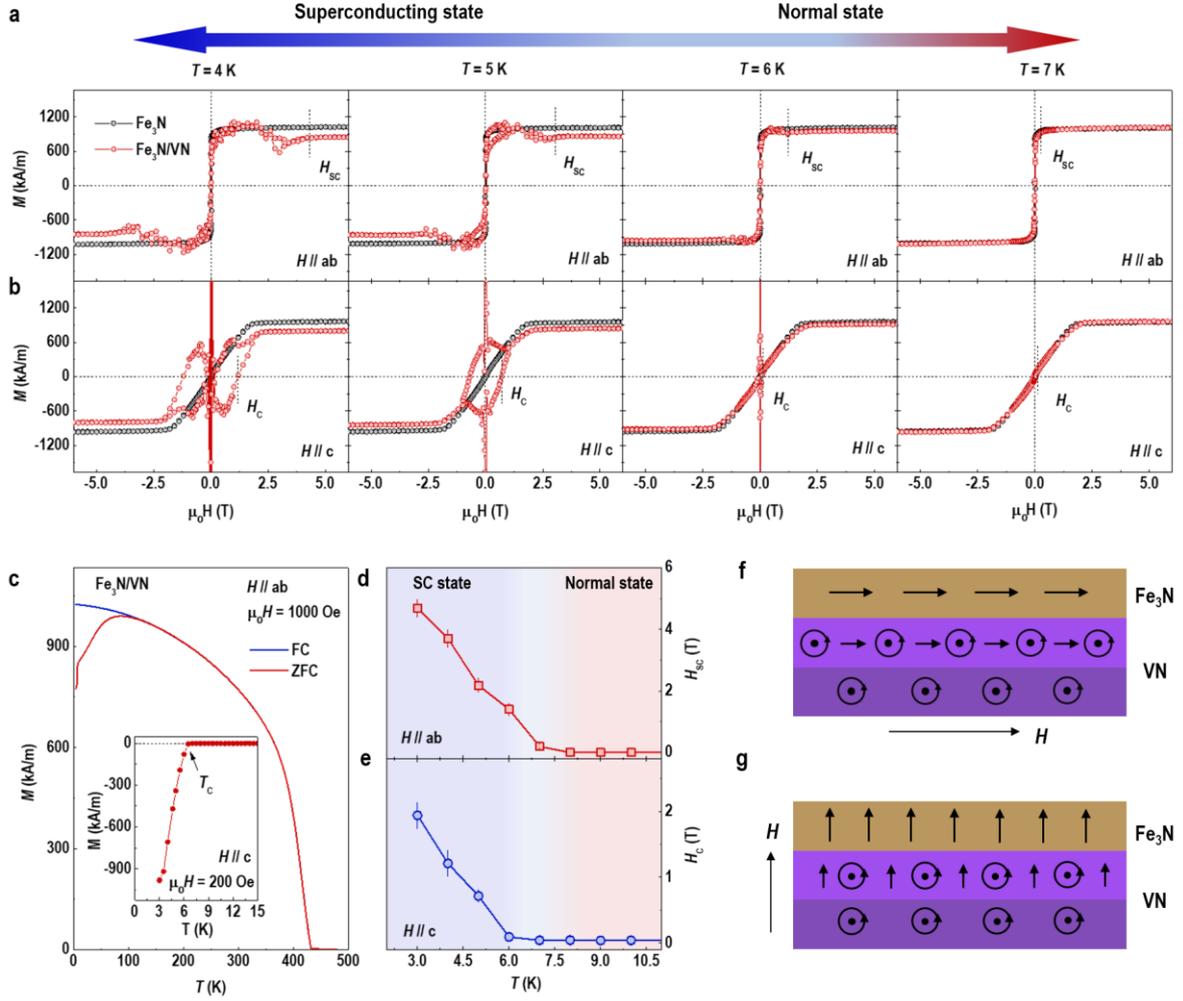

**Figure 3. Strong magnetic coupling at the interfaces between Fe₃N and VN**. (a) and (b) *M-H* hysteresis loops measured from the Fe₃N single layer and Fe₃N/VN bilayer, respectively, with magnetic fields applied parallel to the *ab*-plane and *c*-axis. The temperature for *M-H* loops was varied across the superconducting transition. (c) *M-T* curves of a Fe₃N/VN bilayer. The measurements were carried out during the sample warm-up after zero-field cooling (ZFC) and field cooling (FC) at 1 kOe. The inset of (c) shows the *M-T* curve of a Fe₃N/VN bilayer when an out-of-plane magnetic field of 200 Oe was applied. Temperature dependence of (d) the critical field ($H_{SC}$) and (e) coercive field ($H_C$) under in-plane and out-of-plane magnetic fields. (f) and (g) Illustrations of the spin alignment and magnetic vortices within Fe₃N and VN layers when magnetic fields are parallel or perpendicular to the interfaces, respectively. The illustrations of the vortices are guide for eyes. The orientations of vortices are perpendicular to the applied fields in reality.

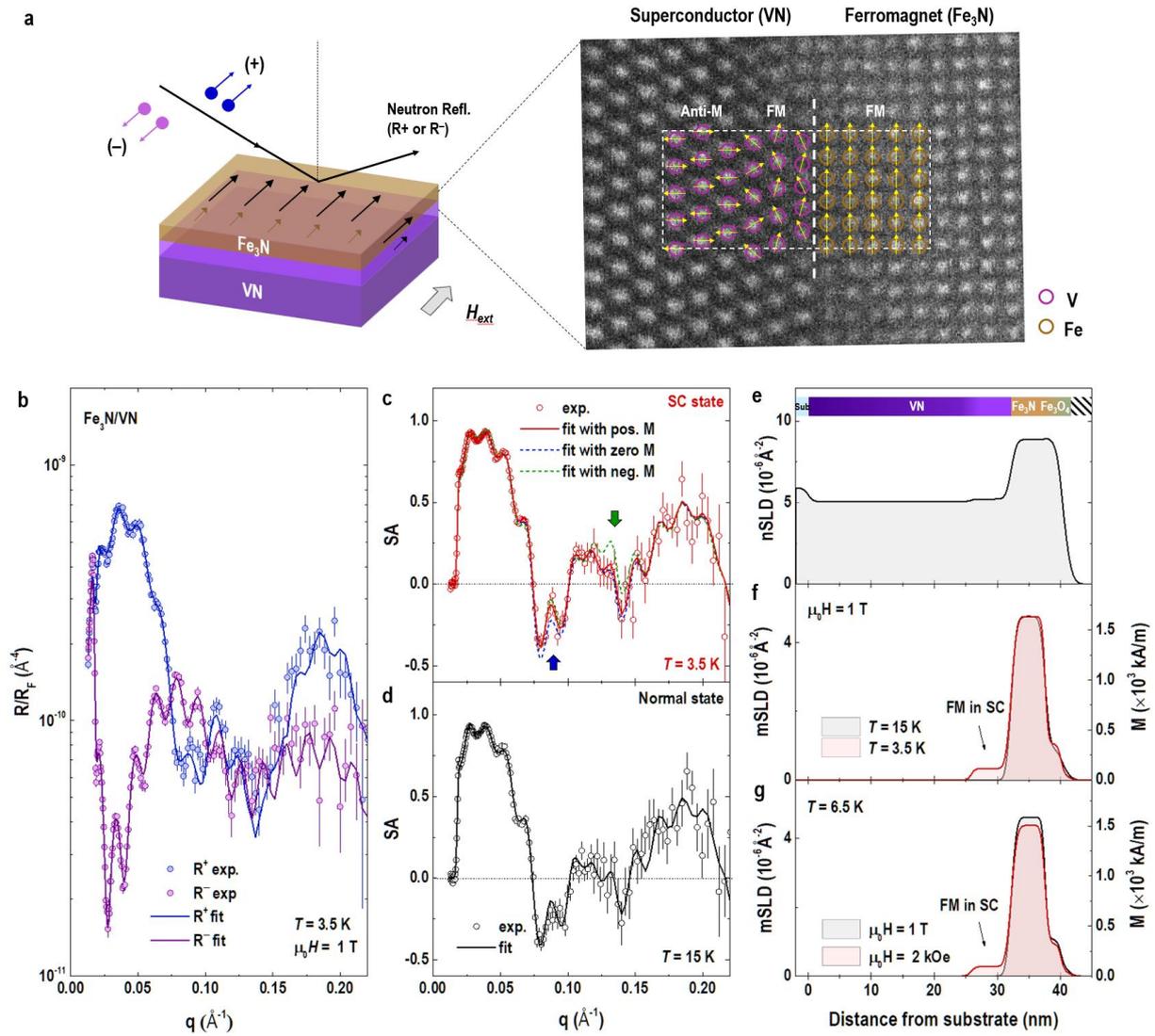

**Figure 4. Magnetic depth profiling across Fe₃N/VN heterostructures**. (a) Schematic of PNR set-up. PNR measurements were performed at fixed temperatures under external in-plane magnetic fields. The inset shows a high-resolution HAADF-STEM image at the interface region of a Fe₃N/VN bilayer, with illustrated spin textures. (b) Measured (open symbols) and fitted (solid lines) reflectivity curves for spin up ($R^+$) and spin down ($R^-$) polarized neutrons as a function of wave vector ($q$). The reflectivities were normalized to the Fresnel reflectivity ($R_F$). PNR spin-asymmetry (SA) ratio SA = ($R^+$+ $R^-$)/( $R^+$–$R^-$) obtained from the experimental and fitted reflectivities when VN in (c) the superconducting (SC) state ($T$ = 3.5 K) and (d) the normal state ($T$ = 15 K) under an external magnetic field of 1 T. The error bars represent one standard deviation. Fits in (c) with zero and negative magnetization in the VN interfacial layer were shown in dashed lines, demonstrating large deviations from experimental data. (e) Nuclear scattering length density (nSLD) and (f) magnetic scattering length density (mSLD) profiles measured for a Fe₃N/VN bilayer at 3.5 and 15 K were presented as a function of the distance from substrate. The scale on the right-hand side shows the absolute magnetization ($M$). (g) PNR mSLD profiles of a Fe₃N/VN bilayer at $T$ = 6.5 K with magnetic fields of 2 kOe and 1 T. The magnetization measured inside VN layer in (f) and (g) is marked with arrows.

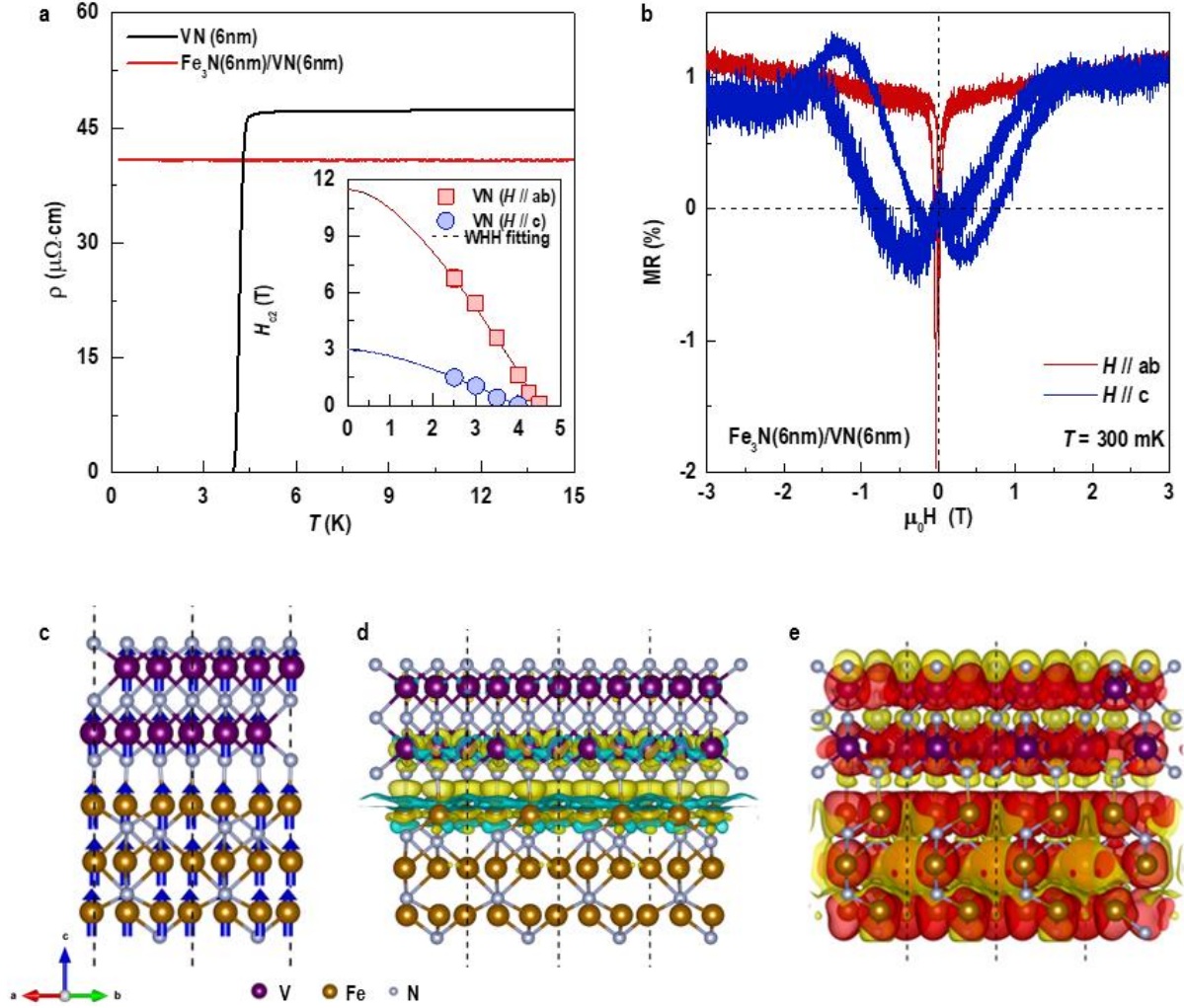

**Figure 5. First-principles calculations on the electronic and spin states across Fe₃N/VN interfaces**. (a) Temperature dependent resistivity of a 6-nm-thick VN single layer and a Fe₃N/VN(6nm) bilayer at zero magnetic field. The bilayer maintains metallic phase down to 300 mK, suggesting the ferromagnetic Fe₃N completely suppressed the superconductivity. Inset shows the temperature dependence of $H_{C2}$ for the 6nm-thick VN single layer. The square and circle symbols represent $H_{C2}$ when the magnetic field was applied parallel to the ab-plane and the c-axis, respectively. The dashed lines are fitting curves based on the WHH model. (b) Field-dependent magnetoresistance (MR) of a Fe₃N/VN(6nm) bilayer at 300 mK. Both in-plane and out-of-plane MR are recorded. (c) The side-view of interface region in a Fe₃N/VN bilayer. The blue arrows indicate the spin orientation of ions. (d) Charge density differential distributions for FM state at the interface when the charge density is equal to $\pm 5\times10^{-3}$ e/Å³. The colored illustrations indicate the isosurfaces, corresponding to the charge accumulation (yellow) and depletion (green) in the space. (e) Contour density corresponding to the FM state for spin up (red) and spin down (yellow) states when the spin density is equal to $\pm 3\times10^{-3}$ μ$_B$/Å³.